\documentclass[seceq]{jpsj3}
\usepackage[usenames]{color}
\def\v#1{\mib #1}
\def\dfrac#1#2{{\displaystyle\frac{#1}{#2}}}

\newcommand{\aver}[1]{\left\langle {#1} \right\rangle}
\newcommand{\avercl}[1]{\left\langle {#1}_{\rm cl}\right\rangle}
\newcommand{\avavcl}[1]{\langle\!\langle {#1}_{\rm cl} \rangle\!\rangle}
\newcommand{\avav}[1]{\langle\!\langle {#1}\rangle\!\rangle}
\def\XHo{\tilde{\Xi}_{\rm odd}}
\def\Nc{N_{\rm c}}

\def\Etot{E^{\rm tot}}
\def\EHal{{\tilde{E}}}
\def\Xcl{\Xi_{\rm cl}}

\def\epsg{{\epsilon_{\rm G}}}
\def\epsgtilde{{\tilde{\epsilon}_{\rm G}}}

\def\ln{{\rm{ln}}}

\def\runauthor{Kazuo {\sc Hida}, Ken'ichi {\sc Takano} and  Hidenori {\sc Suzuki}}
\title
{
Finite Temperature Properties of\\
Mixed Diamond Chain with Spins $1$ and $1/2$\\
}

\author
{
Kazuo {\sc Hida}\thanks{E-mail: hida@phy.saitama-u.ac.jp}, Ken'ichi {\sc Takano}$^{1}$
and  Hidenori {\sc Suzuki}$^{1}$
}

\inst
{Division of Material Science, Graduate School of Science and Engineering, \\ Saitama University, Saitama, Saitama 338-8570\\ 
$^{1}$Toyota Technological Institute, 
Tenpaku-ku, Nagoya 468-851}

\recdate
{
April 8, 2009
}

\abst
{
We formulate statistical mechanics 
for a mixed diamond chain with spins 
1 and 1/2. 
Owing to a series of conservation laws, 
any eigenstate of this system is decomposed into 
eigenstates of finite odd-length spin-1 chains. 
The ground state undergoes five quantum phase transitions 
with varying $\lambda$, a parameter that  controls frustration. 
We obtain the values of the residual entropy and Curie constant which characterize each phase and phase boundary 
at low temperatures. 
We further find various characteristic 
finite-temperature properties such as 
the nonmonotonic temperature dependence of  magnetic susceptibility, 
the multipeak structure in the $\lambda$-dependence of entropy, 
the plateau-like temperature dependence of  entropy and the multipeak structure of specific heat. 
}

\kword{
mixed spin chain, diamond chain, frustration, 
thermodynamics, residual entropy, exact solution} 

\begin{document}

\sloppy

\maketitle

\section{Introduction}
The quantum fluctuation effects of one-dimensional quantum magnets have been extensively studied in various geometrically 
frustrated lattices. 
Owing to the interplay between quantum fluctuation and 
frustration, 
various  exotic quantum phenomena can take place. 
For example, 
spontaneous dimerization,\cite{mg} 1/3-plateau with spontaneous trimerization,\cite{oku,oku2,tone} the transition between quantum and classical plateaus,\cite{ha} 
singlet cluster solid states,\cite{th} 
quantized and partial ferrimagnetisms\cite{
filho,ht,kh}  and spin quadrupolar phases\cite{ht,kh} have been reported. 

Among a variety of models and materials with strong frustration, 
the uniform diamond chain (UDC)  has been attracting the interest of many condensed matter physicists. The UDC consists of successive diamond-shaped units, each unit consisting of spins with equal magnitudes. 
From the theoretical viewpoint, the 
UDC has been rigorously treated to some extent in the absence of distortion.\cite{Takano-K-S} 
The distorted version of this model has also been investigated theoretically.\cite{ottk,otk,sano} 
Surprisingly, it is found that the natural mineral azurite consists of  distorted diamond chains with spin-1/2 and the magnetic properties of this material have been experimentally studied in detail.\cite{kiku,ohta} 
Other materials with the same structure have also been reported.\cite{izuoka,uedia} 

\begin{figure} 
\centerline{\includegraphics[width=6cm]{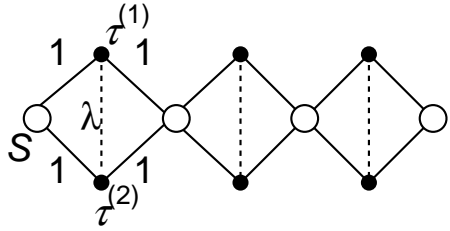}}
\caption{Structure of mixed diamond chain with $S=1$ and $\tau^{(1)}=\tau^{(2)}=1/2$}
\label{lattice}
\end{figure}

In the present work, we  investigate the thermal properties of 
a mixed diamond chain (MDC) with spins 1 and 1/2 
depicted in Fig. \ref{lattice}. 
The MDC has been introduced in a previous paper.\cite{tsh} 
All the eigenstates of this model are represented in terms of the eigenstates of odd-length spin-1 antiferromagnetic Heisenberg chains (AFH1's) and singlet dimers in between. 
The ground state of this model has been rigorously determined in ref. \citen{tsh}, where it is shown that a series of quantum phase transitions take place among phases with different periodicities with spontaneous translational symmetry breakdown. In the present paper, we construct the full thermodynamics of the present model in terms of the eigenstates of the AFH1's with arbitrary odd lengths. In an appropriate parameter regime, only the eigenstates of short AFH1's have a dominant contribution. Therefore, using the numerical diagonalization data for AFH1's, we can obtain reliable numerical results for the thermodynamic properties of infinite mixed diamond chains.  As a result, unique magnetic behaviors are predicted at low temperatures such as the residual entropy and Curie constant which vary from one phase to another. Furthermore, at phase boundaries, these quantities show different behaviors from those in the phases on both sides of the boundary. At higher temperatures, 
 more exotic behaviors are further predicted such as the multipeak structure in the $\lambda$-dependence of entropy, the nonmonotonic temperature dependence of the Curie constant, the multipeak structures of the temperature dependence of specific heat and the plateau-like behavior in the temperature dependence of entropy. Physical interpretations of these behaviors is also presented. 

This paper is organized as follows. In the next section, the model Hamiltonian is presented and the structure of the eigenstates is explained. The ground state properties are summarized in \S 3. Finite temperature statistical mechanics is formulated in \S 4. Low-temperature properties are discussed analytically in \S 5. The numerical analysis of physical properties such as magnetic susceptibility, specific heat  and entropy is presented and discussed in \S 6. The last section is devoted to the summary and discussion. 

\section{Hamiltonian and Cluster Eigenstates} 

The Hamiltonian of the MDC 
in Fig.~\ref{lattice} is represented as 
\begin{align}
{\cal H} &=\sum_{l=1}^{L} J(\v{S}_{l}\v{\tau}^{(1)}_{l}+\v{S}_{l}\v{\tau}^{(2)}_{l}+\v{\tau}^{(1)}_{l}\v{S}_{l+1}+\v{\tau}^{(2)}_{l}\v{S}_{l+1}) + K\v{\tau}^{(1)}_{l}\v{\tau}^{(2)}_{l} , \label{ham}
\end{align}
where $\v{S}_{l}$ and $\v{\tau}^{(\alpha)}_{l}\ (\alpha=1,2)$ 
are spin operators in the $l$th unit cell, 
$J$ and $K$ are the exchange energies, 
and $L$ is the number of unit cells. 
In this paper, we only consider the case in which 
$J$ is positive, and the magnitudes of 
$\v{S}_{l}$ and $\v{\tau}^{(\alpha)}_{l}$
are 1 and 1/2, respectively.  
The ratio $\lambda\equiv K/J$ of the exchange energies 
controls the strength of frustration. 
Below, we set the energy unit $J=1$. 

\begin{figure} 
\centerline{\includegraphics[width=6cm]{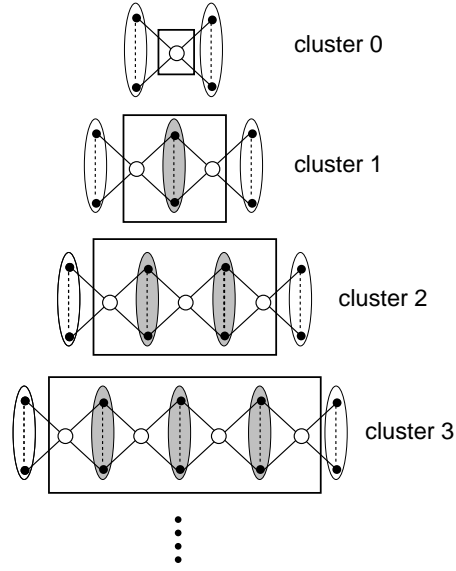}}
\caption{
Series of spin clusters bounded by two singlet dimers. 
A filled (open) circle represents a spin-$1/2$ (spin-1) site. 
An unshaded (shaded) oval represents a singlet (triplet) spin pair. 
A spin cluster is represented by a thick rectangle 
 including triplet pairs  and the spin-1 sites connected to them. 
Cluster-$n$ means a spin cluster including $n$ triplet pairs. 
Each spin cluster has 
 total spin 1 in its lowest-energy state.} 
\label{clusters}
\end{figure}

By defining the composite spin 
$\v{T}_l \equiv \v{\tau}^{(1)}_{l}+\v{\tau}^{(2)}_{l}$ 
for all $l$, Hamiltonian (\ref{ham}) is also expressed as
\begin{align}
{\cal H} &=\sum_{l=1}^{L} \left[\v{S}_{l}\v{T}_{l}+\v{T}_{l}\v{S}_{l+1}+ \frac{\lambda}{2}\left(\v{T}^2_{l}-\frac{3}{2}\right)\right] . 
\label{ham2}
\end{align}
Then, $\v{T}_l^2$ ($l=1,2,..., L$) commutes with ${\cal H}$, 
and $T_l$ defined by $\v{T}_l^2 = T_l(T_l + 1)$ is a good quantum number that takes 0 or 1.  
Therefore, each eigenstate of the Hamiltonian belongs to 
one of the subspaces, each specified by $\{T_l\}$.  
The set $\{T_l\}$ is a sequence of 0's and 1's with the following structure: 
\begin{align}
...0 \, \underbrace{1 \cdots 1}_{n_i} \, 0 \, \underbrace{1 \cdots 1}_{n_{i+1}} \, 0...,
\end{align}
where 
 $n_i$ is the number of 1's bounded by two 0's. 
If $T_l=0$ 
so that $\v{\tau}^{(1)}_{l}$ and $\v{\tau}^{(2)}_{l}$ 
form a singlet dimer state, then we call the spin pair a {\it dimer}. 
The spins on the left of the dimer and those on the right of it 
are rigorously decoupled. 
Therefore, the state of the total spin chain with a specific set $\{T_l\}$ 
is decomposed into dimer states and local spin states between dimers 
in 
a tensor product form. 
We call such a local spin state a spin cluster state, 
and call the assembly of the spins forming the spin cluster state 
the {\it spin cluster}. 
Specifically, we 
call the spin cluster   {\it cluster-$n$} 
if the spin cluster state 
includes $n$ triplet-pair states ($T_l=1$). 
The series of cluster-$n$'s is illustrated in Fig.~\ref{clusters}. 
In particular, two successive 0's imply a single spin-1 site 
bounded by them.  
This spin-1 site is a cluster-0, which is also called a {\it monomer}. 

A cluster-$n$ is equivalent to an AFH1 with length $2n+1$, 
since it consists of $n$ spin-1's for $\v{T}_l $'s and 
$n+1$ spin-1's for $\v{S}_l $'s, as was argued above. 
Thus, we can decompose the total spin chain into AFH1's with lengths $2n_i+1$ with $n_i \geq 0$ 
in each subspace of fixed $\{T_l\}$. 
The lowest-energy state of a cluster-$n$
\cite{name} has total spin 1 from the Lieb-Mattis theorem\cite{lm}. 
Thus, an eigenstate of total Hamiltonian (\ref{ham2}) is specified by a set of quantum numbers $\{n_i, \alpha_i\}$ that satisfy
\begin{align}
\sum_{i=1}^{\Nc } (n_i+1)=L , 
\label{numberofsites}
\end{align}
where the total number of cluster-$n$'s is denoted by $N_{\rm c}$ and $\alpha_i$ specifies the eigenstate of the cluster-$n_i$. 
In the present work, we only consider the thermodynamic quantities in the thermodynamic limit. Therefore, we assume the open boundary condition for simplicity. 

The energy eigenvalue $\Etot$ 
of total Hamiltonian (\ref{ham2}) is 
decomposed into contributions of clusters as 
\begin{align}
\Etot(\{n_i,\alpha_i\},\lambda) = 
\sum_{i=1}^{\Nc} E(n_i,\alpha_i,\lambda) , 
\end{align}
where $E(n,\alpha,\lambda)$ is the $\alpha$-th energy eigenvalue  
of a cluster-$n$. For convenience, we have included the ground state energy of a  neighbouring dimer $-3\lambda/4$ in  $E(n,\alpha,\lambda)$. This can be expressed using the $\alpha$-th eigenvalue $\EHal(2n+1;\alpha)$ of an open AFH1 with length $2n+1$ as
\begin{align}
E(n,\alpha,\lambda)=\EHal(2n+1,\alpha)
+\frac{n\lambda}{4}-\frac{3\lambda}{4}.
\end{align}
 Correspondingly, the lowest-energy  of a cluster-$n$ can be expressed  as
\begin{align}
E_{\rm G}(n,\lambda)=
\EHal_{\rm G}(2n+1) +\frac{n\lambda}{4}-\frac{3\lambda}{4} , 
\end{align}
where $\EHal_{\rm G}(2n+1)$ is the lowest-energy  of AFH1 with length $2n+1$ listed in  Table \ref{table:halg} for $0 \leq n \leq 4$.

\section{Ground States}

The ground states of Hamiltonian (\ref{ham}) are argued in ref.~\citen{tsh} for 
 case of the arbitrary spin magnitude. 
In particular, some rigorous results and their proofs have been given. 
In the case of spins 1 and 1/2, all the ground states and 
quantum phase transitions are determined from numerical 
diagonalization data for finite AFH1's  with odd numbers of spin-1's, 
and the full phase diagram is obtained. 
In this section, we summarize the ground state properties of this model for convenience. 

 The ground state is a uniform array of cluster-$n$'s  with a common value of $n$ and dimers in between. This state is called the dimer-cluster-$n$ (DC$n$) state. The DC0 state is also called the dimer-monomer state. The DC$\infty$ state is the Haldane state. The value of $n$ is determined so as to minimize the total energy for each $\lambda$. This ground state has a spatial periodicity of $n+1$ owing to the $(n+1)$-fold spontaneous translational symmetry breakdown. 
The ground state energy of this DC$n$ phase is given by  
\begin{align}
\Etot_{\rm G}(n,\lambda)=\frac{L}{n+1}E_{\rm G}(n,\lambda).
\end{align}
Accordingly, the ground state energy in the DC$n$ phase per unit cell is given by
\begin{align}
\epsg(n, \lambda)=\frac{1}{n+1}E_{\rm G}(n,\lambda)
=\frac{\EHal_{\rm G}(2n+1)-\lambda}{n+1}+\frac{\lambda}{4}.
\label{ene_unitcell}
\end{align}
If  $T_l=1$ for all diagonal bonds, the ground state energy in the Haldane phase per unit cell is given by
\begin{align}
\epsg(\infty, \lambda)=2\epsgtilde(\infty) +\frac{\lambda}{4},
\end{align}
where $\epsgtilde(\infty)\simeq -1.401484038971$\cite{white} is the ground state energy of an infinite AFH1 per unit cell. 

\begin{table} 
\caption{Ground state energies of odd-length AFH1's.}
\begin{tabular}{|c||c|c|c|c|c|}
\hline
$2n+1$ &1 &  3  &  5  &  7  &  9 \\
\hline
$\EHal_{\rm G}(2n+1)$& 0 & $-3$ & $-5.8302125227708$ & $-8.6345319827062$ &$-11.4329316403302$\\
\hline
\end{tabular}                                                    
\label{table:halg}
\end{table}

\begin{table} 
\caption{Ground state phase  boundaries.}
\begin{tabular}{|c||c|c|c|c|}
\hline
$n_1\ n_2$ &0 \ 1& 1 \ 2  & 2 \ 3  &  3 \ $\infty$ \\
\hline
$\lambda_{\rm c}(n_1,n_2) $& $3$ & $2.660425045542$ & $ 2.58274585704$ &$2.5773403291$\\
\hline                                                           
\end{tabular}                                                    
\label{table:lambdac}
\end{table}

By numerical analysis in our previous study,~\cite{tsh}  
we have found that the DC0, DC1, DC2 and DC3 phases  
appear successively with decreasing $\lambda$ from $+\infty$. 
The phase transition between the DC$(n-1)$ and DC$n$ phases 
takes place at $\lambda=\lambda_{\rm c}(n-1,n)$ given  by
\begin{align}
\lambda_{\rm c}(n-1,n)&=(n+1)\EHal_{\rm G}(2n-1)-n\EHal_{\rm G}(2n+1), 
\label{bdry}
\end{align}
which is the solution of $\epsg(n-1, \lambda)=\epsg(n, \lambda$). 
The critical value is expressed by the ground state energies 
of odd-length AFH1's. 
On the other hand, the DC$n$ phases with $n \ge 4$ do not appear 
with further decrease in $\lambda$, and 
the DC3 phase changes directly into the Haldane phase. 
The phase boundary between the DC3 and Haldane phases 
is given by 
\begin{align}
\lambda_{\rm c}(3,\infty)&=\EHal_{\rm G}(7)-8\epsgtilde(\infty)\simeq 2.5773403291, 
\end{align}
which is the solution of $\epsg(3, \lambda)=\epsg(\infty, \lambda$). 
These critical values of $\lambda$ 
are listed in Table \ref{table:lambdac}.

\section{Formulation of Statistical Mechanics}

\subsection{Grand partition function}

We consider a state that consists of cluster-$n$'s 
with lengths  $n_1, n_2, n_3, \cdots$  and $n_{\Nc }$ 
from left to right. 
For convenience in calculation, we allow  $n_i$ to vary freely. Instead, we introduce the chemical potential $\mu$ conjugate to the number of unit cells in order to fix the expectation value of the total number of  unit cells as eq. (\ref{numberofsites}). 
We also include the external magnetic field $H$ to calculate the magnetic susceptibility $\chi$. 

The grand partition function $\Xi(\beta,\mu,H)$ is given by
\begin{align}
\Xi(\beta,\mu,H) &=\sum_{\Nc =1}^{\infty}
\sum_{\{n_j\}}
\prod_{i=1}^{\Nc } Z_{n_i}(\beta,H) \, e^{\beta\mu(n_i+1)} 
\nonumber\\
&=\sum_{\Nc =1}^{\infty}\left( \sum_{n_i=0}^{\infty}Z_{n_i}(\beta,H) \, e^{\beta\mu(n_i+1)}\right)^{\Nc} , 
\label{grand_partition}
\end{align}
where $\displaystyle \sum_{\{n_j\}}$ stands for 
$\displaystyle \sum_{n_1=0}^{\infty}\sum_{n_2=0}^{\infty}...\sum_{n_{\Nc }=0}^{\infty}$ and 
$Z_n(\beta,H)$ is the partition function of 
a single cluster-$n$ defined by  
\begin{align}
Z_n(\beta,H)&=\sum_{\alpha} 
e^{-\beta (E(n,\alpha,\lambda)-HM(n,\alpha))}
\label{partition_zn}
\end{align}
with the magnetization $M(n,\alpha)$ in the $\alpha$-th eigenstate. 
Equation (\ref{grand_partition}) is then represented as 
\begin{align}
\Xi(\beta,\mu,H) = 
\frac{\Xcl(\beta,\mu,H)}{1-\Xcl(\beta,\mu,H)} 
\end{align}
using the grand partition function $\Xcl(\beta,\mu,H)$ 
of a single cluster defined as 
\begin{align}
\Xcl(\beta,\mu,H)
&=\sum_{n=0}^{\infty} Z_n(\beta,H) \, e^{\beta\mu(n+1)} . 
\end{align}

Suppose that $Q$ is an extensive physical quantity decomposed as
\begin{align}
Q =\sum_{i=1}^{\Nc} Q_{\rm cl}(n_i, \alpha_i), 
\end{align}
then, the grand canonical expectation value $\aver{Q}$ is given by
\begin{align}
\aver{Q} &= \frac{1}{\Xi}{\displaystyle\sum_{\Nc =1}^{\infty}
\sum_{\{n_j\}}\sum_{\{\alpha_j\}}
\left(\sum_{l=1}^{\Nc } Q_{\rm cl}(n_l, \alpha_l)\right)} 
\nonumber\\&\times
\prod_{i=1}^{\Nc } e^{-\beta(E(n_i,\alpha_i,\lambda)-\mu(n_i+1)-M(n_i,\alpha_i)H)} \nonumber\\
&=\frac{\displaystyle\sum_{\Nc =1}^{\infty}{\Nc }\Xcl(\beta,\mu,H)^{\Nc }\aver{Q_{\rm cl}}}{\displaystyle\sum_{\Nc =1}^{\infty}\Xcl(\beta,\mu,H)^{\Nc }} , 
\label{avA0}
\end{align}
where $\avercl{Q}$ is the grand canonical expectation value of 
$Q_{\rm cl}$ in a cluster defined by 
\begin{align}
\avercl{Q}&=  
\frac{1}{\Xcl} \,  
\sum_{n=0}^{\infty}\sum_{\alpha} Q_{\rm cl}(n,\alpha) e^{-\beta(E(n,\alpha,\lambda)-\mu(n+1)-M(n,\alpha)H)} . 
\label{avAcl}
\end{align}
Taking $\Nc = \sum_{l=1}^{\Nc}1$ for $Q$, 
we have $Q_{\rm cl}=1$. 
Then, the expectation value $\aver{\Nc}$ is given as 
\begin{align}
\aver{\Nc }&=\frac{\displaystyle\sum_{\Nc =1}^{\infty}{\Nc }\Xcl(\beta,\mu,H)^{\Nc }}{\displaystyle\sum_{\Nc =1}^{\infty}\Xcl(\beta,\mu,H)^{\Nc }}
=\frac{1}{1-\Xcl}.
\end{align} 
Using $\aver{\Nc }$, eq.~(\ref{avA0}) is written as 
\begin{align}
\aver{Q} = \aver{\Nc }\avercl{Q}. 
\label{avA}
\end{align}
To characterize the size of a cluster-$n$, 
it is convenient to use $L_{\rm cl}\equiv n+1$. 
In fact, the sum of $L_{\rm cl}$'s of all the clusters 
included in the whole chain is the total number of unit cells 
as shown in eq.~(\ref{numberofsites}). 
Taking $L_{\rm cl}$ for $Q_{\rm cl}$ in eq.~(\ref{avAcl}), 
we have 
\begin{align}
\avercl{L}&=  
\frac{1}{\Xcl} \, 
\sum_{n=0}^{\infty} (n+1)\sum_{\alpha} e^{-\beta(E(n,\alpha,\lambda)-\mu(n+1)-M(n,\alpha)H)} . 
\end{align}
Thus, eq.~(\ref{avA}) for $Q=L$ 
 yields 
\begin{align}
\aver{L}&=\aver{\Nc }\avercl{L}=\frac{\avercl{L}}{1-\Xcl} .  
\label{nc}
\end{align}
We define $\avavcl{Q}$ as $\avercl{Q}/\avercl{L}$ for convenience. 
Equations (\ref{avA}) and (\ref{nc}) show that $\avavcl{Q}$ 
is equal to the expectation value per site $\aver{Q}/\aver{L}$ for 
the extensive quantity $Q$. Thus, we have 
\begin{align}
\avavcl{Q} \equiv \frac{\avercl{Q}}{\avercl{L}} 
= \frac{\aver{Q}}{\aver{L}} . 
\label{avform}
\end{align}
Here, we determine the chemical potential $\mu$ in the thermodynamic limit. 
While $\aver{L}$ is a macroscopic quantity, 
$\avercl{L}$ remains always finite at finite temperatures 
because the entropy gain of $O(\ln L)$ overcomes the finite energy cost in 
 making a cut  
in a cluster with  
 length $L$ by virtue of  one-dimensionality. 
Hence, from eq.~(\ref{nc}), 
we must require 
\begin{align}
\Xcl=1 -\frac{\avercl{L}}{\aver{L}}\rightarrow 1\label{mucond}
\end{align}
in the thermodynamic limit. 
This condition fixes $\mu$.

\subsection{Expression in terms of finite-length AFH1}
We can express our formulae using the quantities associated with AFH1. 
For example, the partition function of a cluster-$n$ is expressed as
\begin{align}
Z_n(\beta,H)&= e^{-(n-3)\beta\lambda/4} \, \tilde{Z}(\beta,H;2n+1) , 
\end{align}
where 
\begin{align}
\tilde{Z}(\beta,H;2n+1) &= 
\sum_{\alpha}\exp[-\beta (\EHal(2n+1,\alpha)-HM(n,\alpha))] 
\end{align}
is the partition function of the AFH1 with length $2n+1$ in the magnetic field $H$. Correspondingly, the cluster grand partition function $\Xcl(\beta,\mu,H)$ is expressed as
\begin{align}
\Xcl(\beta,\mu,H)&=
\sum_{n=0}^{\infty}Z_n(\beta,H) \, e^{\beta\mu(n+1)}
= e^{\beta \lambda} \, \XHo(\beta,\tilde{\mu},H) , 
\label{Xcl_XHo}
\end{align}
where $\tilde{\mu}=\mu-\lambda/4$ and $\XHo(\beta,\tilde{\mu})$ is the grand partition function of the odd-length AFH1 given by 
\begin{align}
\XHo(\beta,\tilde{\mu},H)&=\sum_{n=0}^{\infty}\tilde{Z}(\beta,H;2n+1) \, e^{\beta\tilde{\mu} (n+1)}.
\end{align}
Therefore the grand partition function of the whole chain is expressed as
\begin{align}
\Xi(\beta,\mu,H)&=\frac{e^{\beta \lambda} \, \XHo(\beta,\tilde{\mu},H)}{1-e^{\beta \lambda} \, \XHo(\beta,\tilde{\mu},H)} . 
\end{align}
The chemical potential $\tilde{\mu}$ is determined by 
putting $\Xcl = 1$ in eq.~(\ref{Xcl_XHo}) as 
\begin{align}
\lambda&=-T \, \ln\XHo(\beta,\tilde{\mu},H).
\label{murelH}
\end{align}
The cluster expectation value of the physical quantity $Q$ can be written as 
\begin{align}
\avercl{Q}&=\frac{\displaystyle\sum_{n=0}^{\infty}\sum_{\alpha} Q_{\rm cl}(n,\alpha) \, e^{-\beta(\EHal(2n+1,\alpha)-\tilde{\mu} (n+1)-M(n,\alpha)H)}}{\displaystyle\sum_{n=0}^{\infty}\sum_{\alpha} e^{-\beta(\EHal(2n+1,\alpha)-\tilde{\mu} (n+1)-M(n,\alpha)H)}}.
\label{acl}
\end{align}
This expression is useful in numerical calculation, because once the full spectrum of the finite length AFH1 is calculated numerically, we can evaluate the physical quantities of  model (\ref{ham}) using eqs. (\ref{avform}), (\ref{murelH}) and (\ref{acl}) without further massive numerical calculations. 

\subsection{Magnetic susceptibility}
The expectation value of the magnetization $M$ in the magnetic field $H$ is obtained by setting $Q=M$ in eq. (\ref{avform}) as
\begin{align}
\aver{M}&=\aver{L}\avavcl{M} =  
\aver{L}\frac{\avercl{M}}{\avercl{L}} , 
\end{align}
where $\avercl{M}$ is given by eq.~(\ref{avAcl}),  using $M_{\rm cl}$ for $Q_{\rm cl}$. 
Within the first order in $H$, the $H$-dependence of  $\mu$ can be neglected, because $\mu$ is an even function of $H$ from symmetry consideration. Therefore, we find
\begin{align}
\lim_{H \rightarrow 0}\frac{{\avercl{M}}}{H}&=
\lim_{H \rightarrow 0}\frac{\displaystyle\sum_{n=0}^{\infty}\sum_{\alpha} M_{\rm cl}(n,\alpha)^2 \, e^{-\beta(E(n,\alpha,\lambda)-\mu(n+1))}}{T\displaystyle \sum_{n=0}^{\infty}\sum_{\alpha} e^{-\beta(E(n,\alpha,\lambda)-\mu(n+1))}}\nonumber\\
&=\frac{\aver{M_{\rm cl}^2}}{T} , 
\end{align}
where the expectation value $\aver{M_{\rm cl}^2}$ is taken in the absence of the magnetic field. Consequently, the magnetic susceptibility $\chi$ of the total chain is given by
\begin{align}
\chi&=\lim_{H \rightarrow 0}\frac{\aver{M}}{H}=
\frac{\aver{L}}{T}\avav{M_{\rm cl}^2} . 
\end{align}

\subsection{Entropy}
Here and hereafter, we take $H=0$. The entropy $S$ is evaluated using the formula
\begin{align}
S&=
\left.\frac{\partial }{\partial T}T\ln \Xi \right|_{\mu}
=\ln\aver{\Nc }+\frac{\aver{\Nc }}{T}\bigl[\avercl{E}-\mu \avercl{L}\bigr].
\end{align}
In the thermodynamic limit, $\ln \aver{\Nc }$ can be neglected in comparison with $\aver{\Nc }$. Therefore, we find
\begin{align}
S&
=\frac{L}{T} \bigl(\avavcl{E} - \mu \bigr) . 
\label{entform}
\end{align}

\subsection{Internal energy and specific Heat}
The expression for the internal energy $U$ of the total chain is obtained by setting 
$Q(n,\alpha)=E(n,\alpha,\lambda)$ in eq. (\ref{avform}) as
\begin{align}
\frac{U}{L}&=\avavcl{E}= 
\frac{\displaystyle \sum_n\sum_{\alpha} E(n,\alpha,\lambda) \, e^{-\beta (E(n,\alpha,\lambda)-\mu(n+1))}}{\displaystyle \sum_n  (n+1) \sum_{\alpha} e^{-\beta (E(n,\alpha,\lambda)-\mu(n+1))}}.
\label{internal_energy}
\end{align}

Differentiating eq.~(\ref{internal_energy}) 
with respect to $T$ with fixed $L$, 
we have the following expression for specific heat per unit cell:  
\begin{align}
\frac{C}{L} = & 
\frac{1}{T^2}\Biggl[ \avav{E_{\rm cl}^2} - \avavcl{E}\avav{E_{\rm cl}L_{\rm cl}} \Biggr. 
\nonumber \\ 
& - \Biggl. \left.\frac{\partial (\beta\mu)}{\partial \beta}\right|_L  (\avav{E_{\rm cl}L_{\rm cl}} - \avavcl{E}\avav{L_{\rm cl}^2}) \Biggr]. 
\label{spe1}
\end{align}
Since the differentiation of eq.~(\ref{mucond}) for $H=0$ 
with respect to $\beta$ yields 
\begin{align}
\left.\frac{\partial (\beta\mu)}{\partial \beta}\right|_{L} = 
\avavcl{E} , 
\end{align}
eq.~(\ref{spe1}) is written as  
\begin{align}
\frac{C}{L}
=\frac{1}{T^2} \bigl[\avav{E_{\rm cl}^2} - 2\avavcl{E}\avav{E_{\rm cl}L_{\rm cl}} 
+ \avavcl{E}\avav{L_{\rm cl}^2}\bigr] . 
\end{align} 

\section{Low-Temperature Limit}
\subsection{Low-temperature behavior of each phase}

The lowest-energy state of each cluster-$n$ has total spin 1 owing to the Lieb-Mattis theorem\cite{lm}. 
Hence, $L/(n+1)$ spin-1 degrees of freedom remain 
in the DC$n$ phase at $T=0$. 
Accordingly,  
we respectively have the magnetic susceptibility and residual entropy in the DC$n$ phase as 
\begin{align}
\chi &\simeq \frac{2}{3T}\aver{\Nc }=\frac{2}{3(n+1)T}\aver{L},
\label{chi_T_0}\\ 
S &\simeq \aver{\ln 3^{\Nc }}=\frac{\ln3}{n+1}\aver{L}.
\end{align}
Equation (\ref{chi_T_0}) means 
that the Curie constant in each DC$n$ phase is fixed to 
an $n$-dependent value 
(2/3)$L/(n+1)$. 
Therefore, 
we have a stepwise $\lambda$-dependence of $\chi T$,  
as shown by the solid line in Fig. \ref{kvschif}. 
 
\subsection{Low-temperature limit on the phase boundary}
\subsubsection{Magnetic susceptibility}

Near the phase boundary between the DC$(n-1)$ and DC$n$ phases, only the lowest-energy states of cluster-$(n-1)$ and cluster-$n$ contribute to $\XHo$. Therefore, we find
\begin{align}
\XHo(\beta,\tilde{\mu})&=\left[\tilde{Z}(\beta;2n-1)+\tilde{Z}(\beta;2n+1)e^{\beta\tilde{\mu}}\right]e^{\beta\tilde{\mu} n},
\end{align}
where
\begin{align}
\tilde{Z}(\beta;2n\pm 1)&=3\exp(-\beta \EHal_{\rm G}(2n\pm 1)).
\end{align}
According to the formula (\ref{murelH}), 
the chemical potential $\tilde{\mu}$ satisfies the relation 
\begin{align}
\lambda
&=\EHal_{\rm G}(2n-1)-\tilde{\mu} n 
- T \, \ln\left[3 \left(1+e^{-\beta (\EHal_{\rm G}(2n+1)-\EHal_{\rm G}(2n-1)-\tilde{\mu})}\right)\right].
\label{mucr}
\end{align}
On the phase boundary, we combine eqs. (\ref{bdry}) and (\ref{mucr}) to find
\begin{align}
\frac{x}{3}&=\left(\frac{1-x}{x}\right)^{n},\label{aeq}
\end{align}
where
\begin{align}
x=\frac{1}{1+\exp(-\beta (\EHal_{\rm G}(2n+1)-\EHal_{\rm G}(2n-1)-\tilde{\mu})},\label{defA}
\end{align}
which implies that
\begin{align}
\tilde{\mu}=T\ln\frac{1-x}{x}+\EHal_{\rm G}(2n+1)-\EHal_{\rm G}(2n-1).\label{defmu}
\end{align}
The solutions of eq. (\ref{aeq}) are tabulated in the second column of Table \ref{table:xx}. 
For $\lambda=\lambda_{\rm c}(0, 1)$, 
this equation gives an analytic value of $x=(\sqrt{21}-3)/2$ 
($\doteqdot$ 0.791).

\begin{table} 
\caption{$x$, $\chi T/L$ and $S/L$ 
at $\lambda=\lambda_{\rm c}(n-1,n)$ 
in the low-temperature limit.} 
\begin{tabular}{|cc|c|c|c|}
\hline
$n-1$ & $n$ & $x$  & $\chi T/L$ & $S/L$ \\
\hline
0 & 1&  0.791287847477920 & 0.55155122356933 & 1.33270576282026
\\
1 & 2&  0.677814645373914         & 0.28708589748815 &  0.74374685074522 \\
2 & 3&  0.627507207887062         & 0.19767771430847 &  0.52153747001256 \\
\hline                                                           
\end{tabular}                                                    
\label{table:xx}
\end{table}

Using the values of $x$ in Table \ref{table:xx}, the low-temperature susceptibility 
on the phase boundary between the DC$(n-1)$ and DC$n$ phases 
is expressed as
\begin{align}
\chi&= \frac{\aver{L}}{T} \avav{M_{\rm cl}^2} = 
\frac{2\aver{L}}{3T}\frac{{e^{-\beta \EHal_{\rm G}(2n-1)}+e^{-\beta (\EHal_{\rm G}(2n+1)-\tilde{\mu})}}}{ne^{-\beta \EHal_{\rm G}(2n-1)}+(n+1)e^{-\beta (\EHal_{\rm G}(2n+1)-\tilde{\mu})}} \nonumber\\
&=\frac{2\aver{L}}{3T}\frac{1}{n+1-x}.
\label{curiea}
\end{align}
For $\lambda=\lambda_{\rm c}(0, 1)$, this equation gives 
an analytic value of $\chi T/L = (\sqrt{21}+7)/{21}$ 
($\doteqdot$ 0.552). 
in the low-temperature limit. 
The numerical values of $\chi T/L$ 
for all $\lambda=\lambda_{\rm c}(n-1,n)$ 
are tabulated in the third column of Table \ref{table:xx}.

\subsubsection{Entropy}

The entropy $S$ is also expressed by $x$ given by eq. (\ref{aeq}). 
Using eqs.~(\ref{entform}) and (\ref{defmu}), 
 $S$ is calculated as 
\begin{align}
S&=\frac{\aver{L}}{T}
\left[\frac{E_{\rm G}(n-1,\lambda)e^{-\beta \EHal_{\rm G}(2n-1)}+
E_{\rm G}(n,\lambda)e^{-\beta (\EHal_{\rm G}(2n+1)-\tilde{\mu})}}{ne^{-\beta \EHal_{\rm G}(2n-1)}+(n+1)e^{-\beta (\EHal_{\rm G}(2n+1)-\tilde{\mu})}}-\mu\, \right]\nonumber\\
&=\frac{\aver{L}}{T}\frac{(n+1)\EHal_{\rm G}(2n-1)-n\EHal_{\rm G}(2n+1)-\lambda }{n+1-x}-\aver{L}\ln \frac{1-x}{x} , 
\end{align}
Since the first term vanishes as indicated in eq. (\ref{bdry}), 
we finally obtain 
\begin{align}
S=-\aver{L}\ln \frac{1-x}{x}. \label{enta}
\end{align}
The numerically estimated values of $S$ are tabulated in the fourth column of Table \ref{table:xx}.  
For $\lambda=\lambda_{\rm c}(0, 1)$, 
this equation gives an analytic value of 
$S/L = \ln\{(\sqrt{21}+3)/2\}$ ($\doteqdot$ 1.333). 
Note that the entropy at $\lambda=\lambda_{\rm c}(n-1,n)$ is larger than those in the DC$n$ and DC$(n-1)$ phases because of  the excess entropy due to the mixing of two types of clusters. These results can be reproduced by the direct combinatory calculation in the ground state, as shown in Appendix. 

At $\lambda=\lambda_{\rm c}(3,\infty)$,  
 no finite cluster-$n$'s with  $n \ge 4$ can coexist with cluster-3's. Although the excitation energy of  macroscopic cluster-$n$'s with $n \sim O(L)$ vanishes in the thermodynamic limit, 
 the introduction of a macroscopic cluster reduces 
the macroscopic amount of entropy. 
Therefore, no macroscopic clusters can appear at  $\lambda=\lambda_{\rm c}(3,\infty)$ and residual entropy remains the same as that in the DC3 phase.

\section{Numerical Calculation}

\subsection{Estimation of contribution from large clusters}
\label{misent}

We numerically calculate physical quantities 
 using the general formulae derived above. 
Owing to the computer memory limitation, 
we can only include the contribution of cluster-$n$'s 
where the cluster size $n$ is not very large. 
In the following calculations, we denote the largest cluster size $n$ employed in the numerical calculation as $n_{\rm max}$. 

We examine the missing contribution of discarded 
clusters larger than $n_{\rm max}$ by estimating the entropy 
in the high-temperature limit. 
From eq.~ (\ref{entform}), the entropy per unit cell is 
expressed as $S/L = \beta\avavcl{E} - \beta\mu$. 
The first term $\beta\avavcl{E}$ vanishes in the limit of 
$T\rightarrow\infty$, since the physical quantity $\avavcl{E}$ 
is finite. 
Thus, denoting the limiting value of $S/L$ as $S_{\infty}/L$, 
we have 
\begin{align}
\frac{S_{\infty}}{L} = -\lim_{\beta \rightarrow 0} \beta\mu. 
\label{S_infty_L}
\end{align}
We use this relation to estimate the approximate entropy in the high-temperature limit  
formed by cluster-$n$'s with $1 \leq n \leq n_{\rm max}$. 
We denote it as $S_{\infty}(n_{\rm max})$, and 
then $S_{\infty}(\infty)$ is the 
exact value of 
 entropy in the high-temperature limit. 
In this limit, we have $Z_n=\sum_{\alpha}1=3^{2n+1}$ 
from eq.~(\ref{partition_zn}). 
Hence, for $n_{\rm max}$, the cluster grand partition function is 
\begin{align}
\Xcl&=\sum_{n=0}^{n_{\rm max}}3^{2n+1} e^{\beta\mu (n+1)}
= 3e^{\beta\mu}\frac{1 - \left(9e^{\beta\mu}\right)^{n_{\rm max}+1}}{1-9e^{\beta\mu}}. 
\label{Xi_cl_nmax}
\end{align}
From eq.~(\ref{mucond}), we require $\Xcl=1$. 
Then eq.~(\ref{Xi_cl_nmax}) reduces to 
the following equation:
\begin{align}
e^{\beta\mu}&=\frac{1}{12}\left[1+\frac{1}{3}\left(9e^{\beta\mu}\right)^{n_{\rm max}+2}\right] . 
\label{enthi}
\end{align} 
The solution $\beta\mu$ of this equation provides the estimation 
$S_{\infty}(n_{\rm max})/L$ 
for each $n_{\rm max}$. 
The analytic solutions for $n_{\rm max}$ = 0, 1 and 2 
are $S_{\infty}(n_{\rm max})/L =$ $\ln 3$, $\ln [(3/2)(\sqrt{13}+1)]$ and $2 \, \ln 3$, respectively. 
Solutions for $n_{\rm max} \ge 3$ are numerically obtained. 
Also, for $n_{\rm max} \rightarrow \infty$, eq.~(\ref{enthi}) gives $S_{\infty}(\infty)/L=\ln 3+\ln 4$, 
which agrees with  the entropy calculated 
from the number of states 12 for one spin-1 and 
two spin-1/2's per unit cell. 

We present results for $S_{\infty}(n_{\rm max})/L$ 
obtained from eqs.~(\ref{S_infty_L}) and (\ref{enthi}) 
in Table \ref{table:ht}. 
The second column contains the values of 
$S_{\infty}(n_{\rm max})/L$ for $1 \leq n_{\rm max} \leq 5$. 
The third column contains those of the deviations 
$\Delta S_{\infty}(n_{\rm max})/L \equiv$ 
$S_{\infty}(\infty)/L-S_{\infty}(n_{\rm max})/L$. 
For comparison, we show the exact value of 
the high-temperature entropy in the last row. 
This table shows that the missing entropy is only about 3\% 
for $n_{\rm max}=5$, even in the high-temperature limit. 
Therefore, we satisfy ourselves with $n_{\rm max}=5$  
and use only results for $n_{\rm max}=4$ for comparison, if necessary. 

\begin{table} 
\caption{
Contribution of only cluster-$n$'s 
with $0 \le n \le n_{\rm max}$ 
to the entropy in the high-temperature limit. 
The second column contains the approximate entropy 
per unit cell for each $n_{\rm max}$. 
The third column contains the deviation from 
$S_{\infty}(\infty)/L=\ln 3+ \ln 4$, 
which is also shown in the last row. 
}
\begin{tabular}{|c|c|c|c|}
\hline
       $n_{\rm max}$  &  $S_{\infty}(n_{\rm max})/L$ & $\Delta S_{\infty}(n_{\rm max})/L$ \\ 
       \hline
       0  & 1.098612  &   1.386294  \\
       1  & 1.932727  &   0.552179 \\
       2  & 2.197225  &   0.287682 \\
       3  & 2.315806  &    0.169100 \\
       4  & 2.378360  &    0.106547  \\
       5  & 2.414692  &    0.070215 \\
       $\infty$  &   2.484907  &    \\
\hline           
\end{tabular}
\label{table:ht}
\end{table}

\subsection{$\lambda$-dependence of magnetic susceptibility and entropy}

\begin{figure} 
\centerline{\includegraphics[width=6cm]{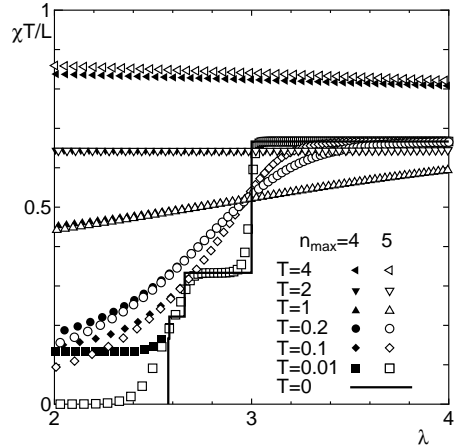}}
\caption{$\lambda $-dependence of $\chi T$ 
with $n_{\rm max}=4$ and 5 
at various temperatures.}
\label{kvschif}
\end{figure}

\begin{figure} 
\centerline{\includegraphics[height=6cm]{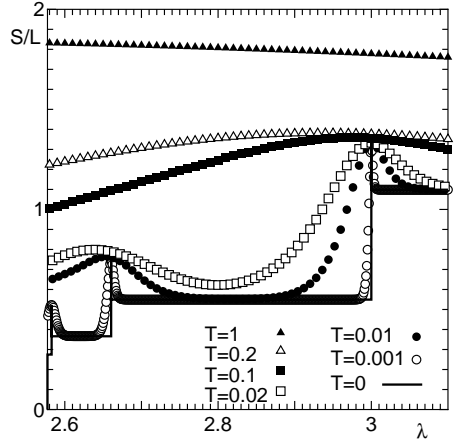}}
\caption{$\lambda$-dependence of entropy with $n_{\rm max}=5$ for $\lambda_{\rm c}(3,\infty) < \lambda <3.1$ 
at various temperatures.}
\label{entlams}
\end{figure}

Figure \ref{kvschif} shows the $\lambda$-dependence of the magnetic susceptibility times the temperature $\chi T$ at various temperatures estimated with $n_{\rm max}=4$ and 5. This quantity approaches the Curie constant in the low-temperature limit.   The stepwise dependence of $\chi T$ in the low-temperature limit is smeared out by thermal fluctuation at finite temperatures. For $\lambda < \lambda_{\rm c}(3,\infty)$, the ground state is the Haldane phase with $n\rightarrow\infty$. Therefore, the present approximation is not reliable. Actually, the result strongly depends on  $n_{\rm max}$(4 or 5) in this region. In contrast, our approximation is fairly reliable for $\lambda >  \lambda_{\rm c}(3,\infty)$, where the ground state is an assembly of finite-size clusters. In this region, the  excited states with large values of $n$  are  unfavorable not only energetically but also  entropically, because the number of such excited states is small, as estimated in \S \ref{misent}.  We actually find that the results in this region are insensitive to $n_{\rm max}$, as shown in Fig. \ref{kvschif}. Therefore, we concentrate on the region $\lambda >\lambda_{\rm c}(3,\infty)$ and fix $n_{\rm max}$ at 5.

The $\lambda$-dependence of entropy is shown in Fig.~\ref{entlams}. 
The peaks due to the excess residual entropy at the phase boundaries are smeared out by the thermal fluctuation with an
increase in temperature. 

\subsection{Cluster excitation energy}

\begin{figure}[b] 
\centerline{\includegraphics[height=6cm]{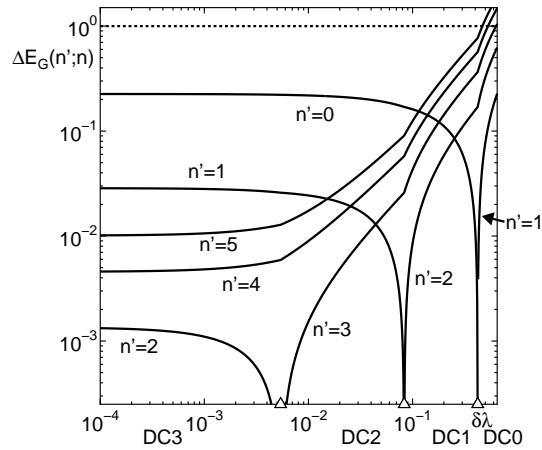}}
\caption{$\lambda$-dependence of the excitation energy $\Delta E_{\rm G}(n';n)$ of cluster-$n'$ in the DC$n$ phase, where $\delta\lambda\equiv\lambda-\lambda_{\rm c}(3,\infty)$. 
The size $n'$ of the excited cluster is indicated for each curve for $0 \leq n' \leq 5$. 
The line $\Delta E_{\rm G}=J$ (=1) is plotted by a dotted line as a  guide for the eye. 
The triangles on the abscissa indicate 
the values of the ground state phase boundaries, and 
each ground state phase is indicated under the abscissa.}
\label{gapcl}
\end{figure}

At finite temperatures, spin clusters other than the cluster-$n$ are excited in the DC$n$ phase. 
One of the simplest excitations is a cluster-$n'$ with $n' \ne n$ 
in its lowest-energy state. 
If a cluster-$n'$ is created in the DC$n$ ground state, 
the system size $L$ changes by $n'-n$. Hence, 
the excitation energy with fixed $L$ 
is given by subtracting the contribution of chemical potential as 
$\Delta E_{\rm G}(n';n)=E_{\rm G}(n',\lambda)-E_{\rm G}(n,\lambda)-(n'-n)\mu$. 
In the low-temperature limit, the condition (\ref{mucond}), 
which determines $\mu$, reduces to the form 
\begin{align}
\Xcl(\beta\rightarrow\infty,\mu)
&= \lim_{\beta\rightarrow\infty} 3e^{-\beta (E_{\rm G}(n,\lambda)-\mu(n+1))}=1,
\end{align}
which implies that $\mu=E_{\rm G}(n)/(n+1)$. 
Thus, the excitation energy required to create 
 a cluster-$n'$ in its lowest-energy state is written as 

\begin{align}
\Delta E_{\rm G}(n';n)
&= (n'+1) \left\{
\frac{\EHal_{\rm G}(2n'+1)-\lambda}{n'+1} 
-\frac{\EHal_{\rm G}(2n+1)-\lambda}{n+1}
\right\} . 
\label{D_E_G}
\end{align}
The $\lambda$-dependence of $\Delta E_{\rm G}(n';n)$ for various values of $n'$ is shown in Fig. \ref{gapcl}, 
where we use 
$\delta \lambda \equiv \lambda-\lambda_{\rm c}(3,\infty)$ 
instead of $\lambda$ itself. 
Although this quantity is 
simply linear in $\lambda$, we employ the log-log plot
to magnify small energy differences.  

\subsection{Temperature dependence of magnetic susceptibility}

\begin{figure}[b] 
\centerline{\includegraphics[height=6cm]{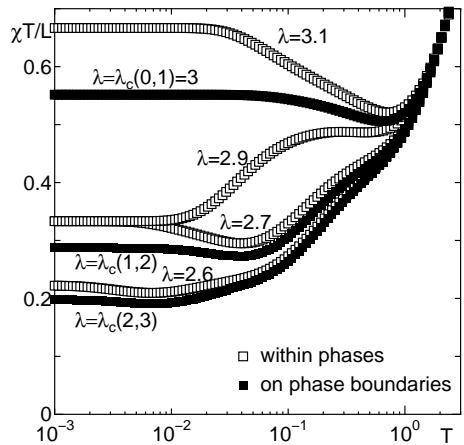}}
\caption{Temperature dependence of $\chi T/L$ within the phases  
(open squares) and on the phase boundaries (filled squares) with $n_{\rm max}=5$.

}
\label{chitr}
\end{figure}

The temperature dependence of  $\chi T$ is presented in Fig. \ref{chitr} for various values of $\lambda$. 
It shows a distinct nonmonotonic behavior.  
By calculating $\chi T$ with $n_{\rm max}$ = 3, 4 and 5, 
we have confirmed that the behavior is insensitive to 
the choice of $n_{\rm max}$. 
Therefore, this is not an artifact of the present approximation.
This behavior is understood by comparing 
Fig.~\ref{chitr} with Fig.~\ref{gapcl} as follows. 

We investigate  
$\chi T$ near each phase boundary $\lambda_{\rm c}(n-1,n)$. 
If $\lambda$ approaches  $\lambda_{\rm c}(n-1,n)$ 
in the DC$n$ phase ($\lambda < \lambda_{\rm c}(n-1,n)$), 
the excitation energy $\Delta E_{\rm G}(n-1;n)$ of cluster-$(n-1)$ comes down to zero, as shown in Fig.~\ref{gapcl}. 
Hence, cluster-$(n-1)$'s mix with cluster-$n$'s at low temperatures. 
A cluster-$(n-1)$ has the same spin unity as a cluster-$n$, and 
the cluster length of a cluster-$(n-1)$ is shorter by a unit. 
Accordingly, such excitations increase the density of alive spins,
 which lead to an increase in $\chi T$ at low temperatures. 
This behavior is seen, e.g., for 
$\lambda$ = 2.9 near $\lambda_{\rm c}(0, 1)$ = 3 
in $10^{-2} \lesssim T \lesssim 10^{-1}$ in Fig.~\ref{chitr}. 
On the other hand, if $\lambda$ approaches  $\lambda_{\rm c}(n-1,n)$ 
in the DC$(n-1)$ phase ($\lambda > \lambda_{\rm c}(n-1,n)$), 
the excitation energy $\Delta E_{\rm G}(n;n-1)$ of cluster-$n$ comes down to zero, as shown in Fig.~\ref{gapcl}. 
Hence cluster-$n$'s mix with cluster-$(n-1)$'s at low temperatures. 
Such excitations decrease the density of alive spins, and lead to the decrease in $\chi T$. 
This behavior is seen, e.g., for 
$\lambda$ = 2.7 near $\lambda_{\rm c}(1, 2)$ $\doteqdot$ 2.660 in $10^{-2} \lesssim T \lesssim 0.5 \times 10^{-1}$ in Fig.~\ref{chitr}.

On the phase boundary of $\lambda =\lambda_{\rm c}(n-1,n)$, 
the energies per unit cell of  cluster-$(n-1)$ and cluster-$n$ degenerate. 
In this case, the next-lowest excitation  
controls the temperature dependence of $\chi T$. 
At $\lambda=\lambda_{\rm c}(n,n-1)$, the candidates for the lowest excitation energy are $\Delta E_{\rm G}(n-2;n)$ and  $\Delta E_{\rm G}(n+1;n)$. 
Using eqs. (\ref{bdry}) and (\ref{D_E_G}), 
these are written as 
\begin{align}
\Delta E_{\rm G}(n-2;n)&=\EHal_{\rm G}(2n+1)+\EHal_{\rm G}(2n-3) - 2\EHal_{\rm G}(2n-1),\\
\Delta E_{\rm G}(n+1;n)&=\EHal_{\rm G}(2n+3)+\EHal_{\rm G}(2n-1)-2\EHal_{\rm G}(2n+1). 
\end{align} 
Because of their second difference forms, 
they are independent of the bulk term proportional to $n$ and 
the constant boundary term in $\EHal_{\rm G}(2n+1)$. 
Accordingly, $\Delta E_{\rm G}(n-2;n)$ and 
$\Delta E_{\rm G}(n+1;n)$ are only determined by the 
$n$-dependent boundary term resulting 
from the interaction between the boundary spins of 
the cluster-$n$. 
Since this interaction decreases with $n$, 
$\Delta E_{\rm G}(n+1;n)$ is lower than $\Delta E_{\rm G}(n-2;n)$. 
Figure \ref{gapcl} actually shows that, at each phase boundary 
$\lambda_{\rm c}(n-1,n)$, 
the next-lowest cluster is cluster-$(n+1)$ 
with an excitation energy of $\Delta E_{\rm G}(n+1;n)$. 
The mixing of cluster-$(n+1)$'s into cluster-$(n-1)$'s and 
cluster-$n$'s decreases the density of alive spins. 
Therefore, even at the phase boundary, $\chi T$ decreases with $T$ at low temperatures. 

At higher temperatures, 
clusters with higher energies affect $\chi T$. 
Figure \ref{gapcl} shows that 
cluster-$(n-2)$'s come into play 
for $\lambda \geq \lambda_{\rm c}(n-1,n)$, 
e.g. an excitation of $n'$ = 0 appears 
in DC$1$ $(\lambda > \lambda_{\rm c}(1, 2))$ phase and 
at the phase boundary  $\lambda = \lambda_{\rm c}(1, 2)$. 
The mixing of these shorter clusters increases $\chi T$, 
which is decreased by cluster-$(n+1)$'s at low temperatures.

At a general temperature less than $J$, short and long clusters 
with various excitation energies contribute to $\chi T$, and 
the shorter ones increase and the longer ones decrease $\chi T$. 
Which clusters mainly contribute to $\chi T$ depends on 
temperature, and this qualitatively explains 
the nonmonotonic  temperature dependence of $\chi T$. 
For $T \gtrsim J$, $\chi T$ increases monotonically toward the free-spin limit $\chi T/L=2/3+1/2=7/6$. 
The temperature dependence of $\chi T$ is insensitive to $\lambda$ in this regime.

\subsection{Temperature dependence of specific heat and entropy}

\begin{figure} 
\centerline{\includegraphics[width=6cm]{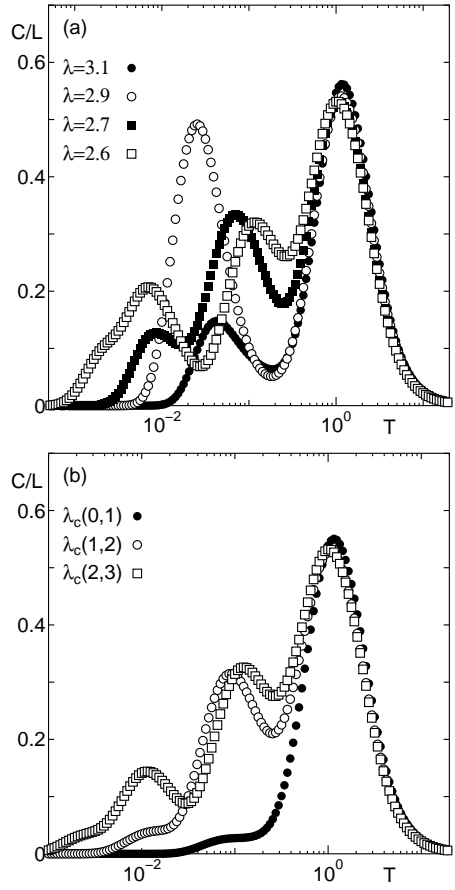}}
\caption{Temperature dependence of  specific heat (a) in each phase  and (b) on each phase boundary 
by  calculation with $n_{\rm max}=5$.}
\label{hine}
\end{figure}

\begin{figure} 
\centerline{\includegraphics[height=6cm]{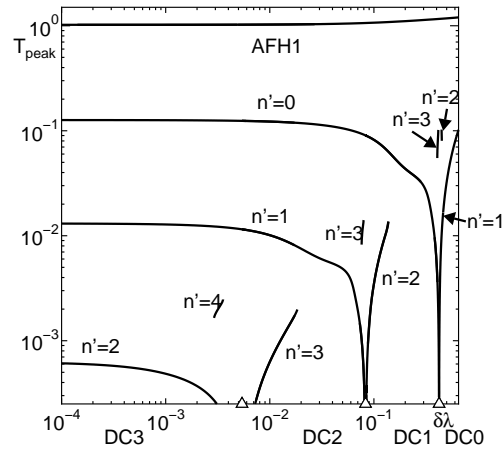}}
\caption{$\lambda$-dependence of the peak temperature $T_{\rm peak}$ of  specific heat where $\delta\lambda\equiv\lambda-\lambda_{\rm c}(3,\infty)$. The size $n'$ of the cluster responsible for each peak is indicated for each curve with $n_{\rm max}=5$. The ground state phases are indicated below the abscissa. The triangles are the ground state phase boundaries.}
\label{peaks}
\end{figure}

\begin{figure} 
\centerline{\includegraphics[width=6cm]{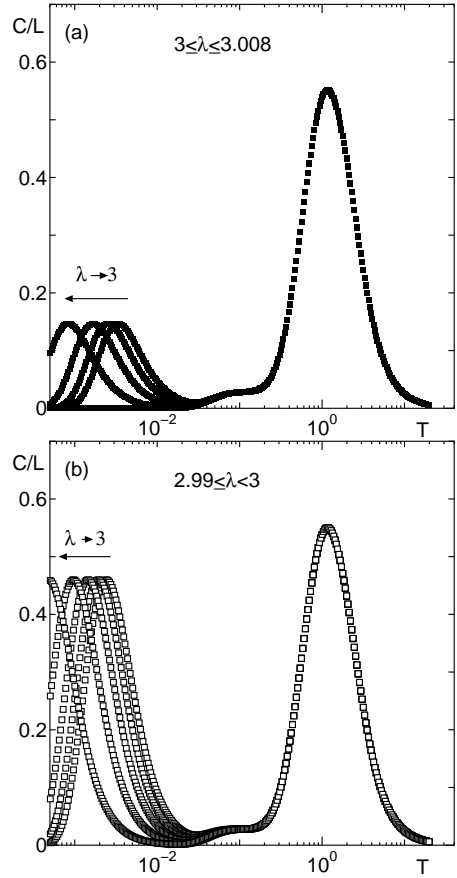}}
\caption{Temperature dependence of specific heat around the phase boundary of $\lambda_{\rm c}(0, 1) = 3$ with $n_{\rm max}=5$ for (a) $\lambda  \geq 3$ and (b)  $\lambda  < 3$.}
\label{hinecr2}
\end{figure}

\begin{figure}[!] 
\centerline{\includegraphics[width=6cm]{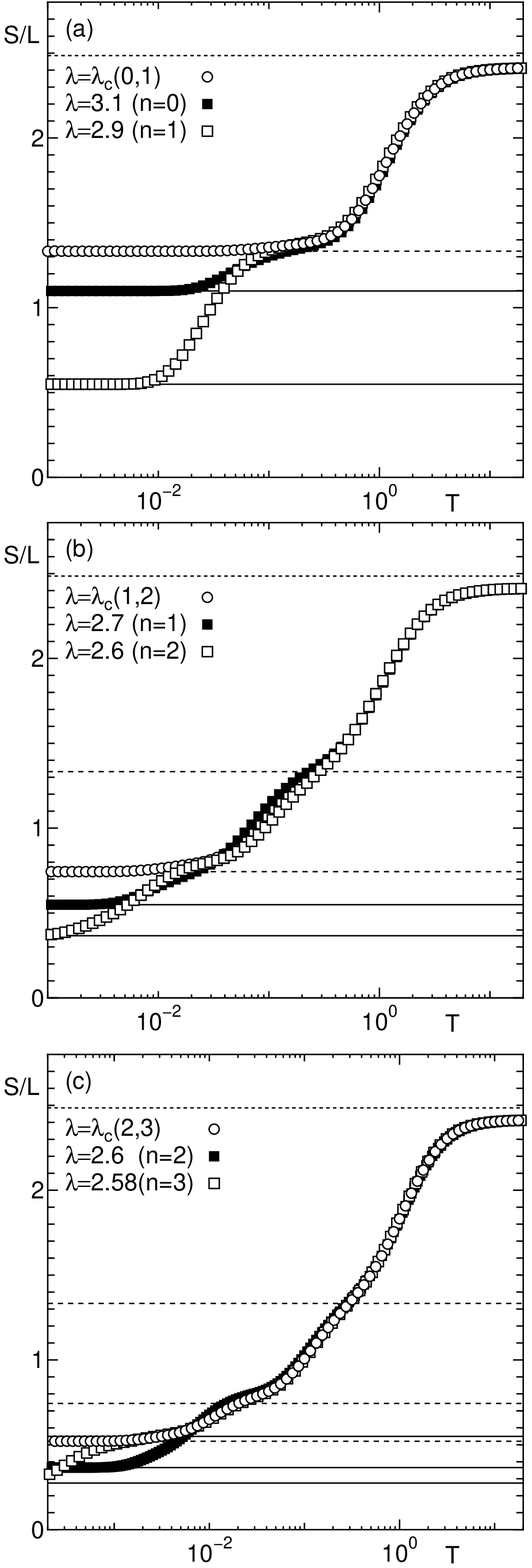}}
\caption{Temperature dependence of entropy  around (a) $\lambda=\lambda_{\rm c}(0,1)$, (b) $\lambda=\lambda_{\rm c}(1,2)$ and (c) $\lambda=\lambda_{\rm c}(2,3)$ with $n_{\rm max}=5$. The open circles, filled squares and open squares represent the values for $\lambda=\lambda_{\rm c}$,  $\lambda > \lambda_{\rm c}$ and  $\lambda < \lambda_{\rm c}$, respectively. The dotted line indicates the correct value of entropy in the high-temperature limit. The solid lines are the values of residual entropy in each phase. The broken lines are the values of residual entropy on the phase boundaries.}
\label{entg}
\end{figure}

The temperature dependence of  specific heat is presented in Fig. \ref{hine}. In general, specific heat shows a multipeak structure. This structure is 
also understood  on the basis of the excitation spectrum $\Delta E_{\rm G}(n';n)$.

The $\lambda$-dependence of the peak position 
is shown in Fig. \ref{peaks}. 
The uppermost peak is common to all values of $\lambda$, 
and originates from the bulk excitations of cluster-$n$'s, which are equivalent to finite AFH1's. 
The positions of the lower peaks strongly depend on $\lambda$. 
When $\lambda$ approaches a phase boundary from any side, a peak approaches $T=0$, as shown in Fig. \ref{peaks}. 
By comparing  Fig. \ref{peaks} with Fig. \ref{gapcl}, 
such a peak is interpreted to originate from the excitation 
of a cluster-$(n+ 1)$ or cluster-$(n- 1)$ with an excitation energy of 
$\Delta E_{\rm G}(n \pm 1;n)$ in the DC$n$ ground state. 
Approaching the phase boundary, 
$\Delta E_{\rm G}(n+1;n)$ comes down to zero for $\lambda \geq \lambda_{\rm c}(n,n+1)$ 
and $\Delta E_{\rm G}(n-1;n)$ comes down to zero for $\lambda \leq \lambda_{\rm c}(n-1,n)$.
Hence, at the phase boundary, cluster-$n$'s and cluster-$(n-1)$'s  mix, and yield 
the excess residual entropy at $T=0$. 
We show the temperature dependence of  specific heat around the phase boundary of $\lambda_{\rm c}(0, 1) = 3$ in Fig. \ref{hinecr2}, where specific heats are plotted at 
$\lambda =2.99$ to 3.008 with an interval of 0.002. 
The lower peak shifts toward $T=0$ 
as $\lambda $ approaches $\lambda_{\rm c}(0,1)=3$ from both above and below. At $\lambda=3$, the lower peak vanishes. 
Also, in the neighbourhood of other phase boundaries, specific heat behaves similarly.
 
Entropy is estimated using the formula (\ref{entform}). 
The temperature dependence of the entropy is shown in Fig. \ref{entg}, 
where panels (a), (b) and (c) are for regimes around 
$\lambda=\lambda_{\rm c}(0,1)$, $\lambda=\lambda_{\rm c}(1,2)$ and $\lambda=\lambda_{\rm c}(2,3)$, respectively. 
We find that entropy has a plateau-like behavior 
around the value of the residual entropy on phase boundaries. 
This behavior is explained  
as  follows. 

Between the temperatures corresponding to $J$ 
 and $\Delta E_{\rm G}(0;1)$ ($\lambda < \lambda_{\rm c}(0,1)$) or $\Delta E_{\rm G}(1;0)$ ($\lambda > \lambda_{\rm c}(0,1)$), there exists a temperature regime where bulk excitations of the finite-length AFH1 are not excited 
 and several cluster-$n$'s with  $n \geq 0$, which have lower excitation energies than the Haldane gap, are still equally excited in their lowest-energy states. 
In this regime, the entropy 
 is close to the residual entropy at $\lambda=\lambda_{\rm c}(0,1)$ because the contribution of cluster-$n$'s with $n \ge 2$ is small. For large $\lambda \gg  \lambda_{\rm c}(1,2)$, this behavior continues until the true ground state is chosen at lower temperatures, as shown in Fig. \ref{entg}(a). With a decrease in $\lambda$, however, $\Delta E_{\rm G}(0;1)$ increases while $\Delta E_{\rm G}(n;1)$'s with $n \ge 2$ decrease. Therefore, there exists a temperature regime where  cluster-$n$'s with $n\geq 1$, which have lower $\Delta E_{\rm G}(n;1)$ than $\Delta E_{\rm G}(0;1)$, are still equally excited in their lowest-energy states. The entropy of this state is close to the residual entropy at $\lambda=\lambda_{\rm c}(1,2)$,  as shown in Fig. \ref{entg}(b), because the contribution of cluster-$n$'s with $n \ge 3$ is small. Similarly, the temperature dependence of entropy has plateau-like structures around the value of the residual entropy on the phase boundary at $\lambda_{\rm c}(k-1,k)$, if $\lambda \simeq\lambda_{\rm c}(n-1,n)$ with $n \geq k$.

\section{Summary and Discussion}

The statistical mechanics of the MDC with 
spins 1 and 1/2
is formulated rigorously. 
The low-temperature behaviors of 
the magnetic susceptibility, specific heat and entropy 
are analytically investigated.  
The Curie constant and residual entropy vary depending on the ground state phase. 
At finite temperatures, these quantities are 
obtained  
using the exact numerical diagonalization data 
for odd-length AFH1's. 
The finite-temperature behavior of physical quantities also varies reflecting the ground state properties and excitation spectra. Magnetic susceptibility shows a nonmonotonic temperature dependence. Specific heat has a multipeak structure. 
At low temperatures, entropy as a function of $\lambda$  
has peaks around the ground state phase boundaries reflecting the ground state excess entropy. 
Entropy as a function of temperature exhibits 
plateau-like structures depending on  $\lambda$. 
An intuitive explanation for these exotic features based on the cluster picture is given. 
Similar behaviors are also found in the frustrated Ising model on a diamond hierarchical lattice.\cite{fuku}

Although no real material described by 
the MDC model has been known so far, 
the MDC with distortion will  hopefully be realized. 
It is well known  that 
the natural mineral azurite, in which Cu ions carry spin-1/2 degrees of freedom\cite{kiku}, is a distorted version of the spin-1/2 UDC. 
Another material for the 
distorted UDC
 with spin-1/2 is also known.\cite{uedia} 
These facts encourage the search for materials described by the distorted version of the MDC.
As long as the distortion is weak, the effect of distortion is smeared out at finite temperatures and the phenomena predicted in the present work would be observable. 
On the other hand, the ground state properties are sensitive to lattice distortion because 
the DC$n$ ground states of the present model have a macroscopic number of effectively free spin-1 degrees of freedom. 
  A variety of quantum phases can emerge in the presence of perturbation. Therefore, it is also important to investigate the effect of lattice distortion on the ground state of  MDC's theoretically.   
Our preliminary study suggests the presence of a variety of exotic phases and phenomena, such as quantized and partial ferrimagnetic phases, Haldane phases with broken translational symmetry and an infinite series of quantum phase transitions depending on the type of distortion. 
Further investigation of these phenomena is in progress and will be reported in future publications.

The interchain interaction is another important perturbation 
that controls the low-temperature behavior of the MDC, 
if it is realized as a real material. 
It is naturally expected that the spin-1 degrees of freedom, which remain paramagnetic in the ground state of a single chain, become ordered in the presence of an interchain interaction. 
The competition of magnetic ordering against 
the various phenomena reported in this work 
is also an interesting issue. 

The numerical diagonalization program is based on the package TITPACK ver.2 coded by H. Nishimori.  The numerical computation in this work has been carried out using the facilities of the Supercomputer Center, Institute for Solid State Physics, University of Tokyo and the Supercomputing Division, Information Technology Center, University of Tokyo.  This work is  supported by a Grant-in-Aid for Scientific Research  on Priority Areas, "Novel States of Matter Induced by Frustration", from the Ministry of Education, Science, Sports and Culture of Japan, and  
Fund for Project Research in Toyota Technological Institute. 

\appendix
\section{Combinatorial derivation of the residual entropy}

We reproduce the entropy in \S 5.2.2 
by direct combinatory calculation. 
On the phase boundary between the DC$(n-1)$ and DC$n$ phases, two kinds of clusters, cluster-$(n-1)$ and cluster-$n$, coexist. 
We denote the number of cluster-$(n-1)$'s by $N_{n-1}$ and that of cluster-$n$'s by $N_n$. 
Under the constraints $(n-1)N_{n-1}+nN_{n}=L$ and $N_{n-1}+N_{n}=\Nc $, the total number of the allowed states is given by 
\begin{align}
W&=\frac{(N_{n-1}+N_{n})!}{N_{n-1}!N_{n}!}3^{N_{n-1}+N_{n}}\nonumber\\
&\simeq\frac{(N_{n-1}+N_{n})^{N_{n-1}+N_{n}}}{N_{n-1}^{N_{n-1}}N_{n}^{N_{n}}}3^{N_{n-1}+N_{n}}
\label{comb}
\end{align}
in the thermodynamic limit. 
Here, we have taken into account the 3-fold degeneracy of 
the spin-1 lowest-energy state for each cluster. 
Setting $\Nc =Lc$, $N_{n-1}=\Nc x$ and $N_{n}=\Nc (1-x)$, 
we have $c=\dfrac{1}{n+1-x}$. 
Then, eq.~(\ref{comb}) is written as 
\begin{align}
W&=\frac{1}{x^{Lcx}(1-x)^{Lc(1-x)}}3^{Lc}, 
\end{align}
and entropy is calculated as 
\begin{align}
S&=\ln W 
=-\frac{L}{n+1-x}\left[x\ln x +(1-x)\ln(1-x)-{\ln 3}\right].
\label{entrop}
\end{align}
Optimizing $S$ with respect to $x$, we have
\begin{align}
\frac{\partial S}{\partial x}&=
-\frac{L}{(n+1-x)^2}\left[x\ln x +(1-x)\ln(1-x)-{\ln 3}\right]\nonumber\\
&-\frac{L}{n+1-x}\ln \frac{x}{1-x}=0.
\end{align}
This equation is equivalent to eq.~(\ref{aeq}) determining $x$, 
and eq.~(\ref{entrop}) reduces to the expression (\ref{enta}) 
for  residual entropy.


\begin{thebibliography}{50}
\bibitem{name} We use the word 'lowest-energy state' for cluster-$n$ instead of 'ground state' to avoid confusion with the ground state of the whole MDC.
\end{thebibliography}
\end{document}